# Spalling-induced liftoff and transfer of electronic films using a van der Waals release layer

*Eric W. Blanton\*, Michael J. Motala, Timothy A. Prusnick, Albert Hilton, Jeff L Brown, Arkka Bhattacharyya, Sriram Krishnamoorthy, Kevin Leedy, Nicholas R. Glavin, Michael Snure*

E. W. Blanton, T. A. Prusnick, A. Hilton, J. L. Brown
KBR, Beavercreek, OH 45431, USA
E-mail: eric.blanton.5.ctr@us.af.mil

M. J. Motala
UES, Dayton, OH 45432, USA

A. Bhattacharyya, S. Krishnamoorthy
University of Utah, Electrical and Computer Engineering, Salt Lake City, UT 84112, USA

K. Leedy, M. Snure
Air Force Research Laboratory, Sensors Directorate, WPAFB, OH 45433, USA

N. R. Glavin
Air Force Research Laboratory, Materials and Manufacturing Directorate, WPAFB, 45433, USA

Heterogeneous integration strategies are increasingly being employed to achieve more compact and capable electronics systems for multiple applications including space, electric vehicles, and wearable and medical devices. To enable new integration strategies, the growth and transfer of thin electronic films and devices, including III-nitrides, metal oxides, and two-dimensional (2D) materials, using 2D boron nitride (BN)-on-sapphire templates is demonstrated. The van der Waals BN layer, in this case, acts as a preferred mechanical release layer for precise separation at the substrate-film interface and leaves a smooth surface suitable for van der Waals bonding. A tensilely-stressed Ni layer sputtered on top of the film induces controlled spalling fracture which propagates at the BN/sapphire interface. By incorporating controlled spalling, the process yield and sensitivity is greatly improved, owed to the greater fracture energy provided by the stressed metal layer relative to a soft tape or rubber stamp. With stress playing a critical role in this process, the influence of residual stress on detrimental cracking and bowing is investigated. Additionally, a selected area lift-off technique is developed which allows for isolation and transfer of individual devices while maximizing wafer area use and minimizing extra alignment steps in the integration process.

## 1. Introduction

With the demand for increased connectivity for 5G communications and the internet of things (IoT), high performance electronic systems are being incorporated in new ways which require multiple components, functions, and technologies to be combined in a single chip or package.[1] Heterogeneous integration at the chiplet level has proven successful at improving system power, performance, and area (PPA) scaling while decreasing design time and fabrication costs.[1] Achieving even closer integration of the components by isolating just the thin active material layers from the growth substrate can further improve the size, weight, and power (SWAP) by providing more opportunities for optimization of the dielectric material or heat spreading layer in contact with the device.[2-3] Additionally, for space or wearable applications where small size is a priority, the very thinness of the components can enable mechanically flexible or conformal form factors not possible using chiplets.[1,4]

One of the key challenges to achieving this closer integration is isolating the active material layer from the growth substrate, which is typically performed using variations of the following methods.[5] Epitaxial lift-off (ELO) strategies use a wet etch which undercuts the epitaxial layer at a sacrificial insert layer and are typically used for thin films such as GaAs and InP,[6-8] though a photoelectrochemical variant of the method has been used with GaN.[9] Because a specialized sacrificial insert layer is required, this method cannot be easily adapted to many material systems. Laser liftoff is typically used with GaN-on-sapphire and relies on the absorption difference between the GaN and substrate to decompose a thin region of GaN near the substrate interface.[10] Since the substrate bandgap must be higher than GaN for this method to work, GaN-on-SiC processes cannot utilize the laser liftoff technique. Laser liftoff also typically suffers from rough surfaces after separation which limit the options for secondary bonding.[11] The van der Waals liftoff method can be used to separate membranes from the growth substrate using moderate mechanical force.[12] In this method, a two-dimensional material is embedded in a layer stack and its weak out-of-plane bonding allows for separation using tape or polydimethylsiloxane (PDMS) stamping. Although fabricated devices have been separated using this method,[13-17] it is challenging to achieve 2D layer adhesion that is robust enough to survive processing conditions without spontaneously delaminating while still being weak enough to reliably separate using soft adhesive transfer methods.[18-21] Controlled spalling is another layer exfoliation method which involves depositing a tensilely stressed layer on the material to be separated, and the resulting forces cause a crack to propagate in the substrate parallel to the surface.[22-23] In a homogeneous material, the depth of the fracture is determined by the thickness and stress of the deposited layer, so any inhomogeneity in stressor layer properties will cause the fracture to deviate upwards or downwards in the material.[24] Additionally, the feasibility of acquiring a high-quality fracture interface is dependent on the crystal structure and orientation of the substrate.[25]

By using controlled spalling with an inserted van der Waals layer we can take advantage of the best qualities from each method. The embedded van der Waals layer makes the fracture surface uniformly flat by acting as a preferential fracture layer.[26] At the same time, the larger fracture energy of the deposited stressor layer allows us to use a better-adhered release layer which will not spontaneously delaminate during device processing. The damage is limited to a couple of monolayers of material, enabling very thin layers to be transferred, and the surfaces have sub-nanometer roughness, in contrast to the surfaces attained from laser liftoff and epitaxial liftoff. This idea has been used to transfer membranes of several technologically important materials including III-V semiconductors, graphene and other 2D materials, and complex oxide materials.[27-29] Here we use 2D BN grown on sapphire as a template and extend the versatility of this method by exfoliating and transferring processed devices and a variety of electronic films. We demonstrate the separation and transfer of AlGaN/GaN high electron-mobility transistors (HEMTs) as well as other thin films including the BN layer itself, AlN, $Ga_2O_3$, ZnO and transition metal dichalcogenide (TMD) thin films using the vdW-mediated controlled spalling method. Additionally, a post growth/processing selective area transfer process is presented which will enable pick-and-place transfer of individual devices and patterned arrays of devices.

## 2. Results and Discussion

The exfoliation and transfer process using BN-on-sapphire templates is depicted in **Figure 1a**. After the BN is grown, the film to be transferred is grown or deposited on the BN-on-sapphire template[30]. The versatility of the BN-on-sapphire template is demonstrated by depositing a variety of electronic materials including oxides, nitrides, and TMDs, by a range of deposition methods including pulsed laser deposition (PLD) and metal organic chemical vapor deposition (MOCVD). Following film deposition, a 1 to 6 µm thick film of tensilely stressed Ni is sputtered on top. To initiate spalling and resultant separation of the thin film, a crack is formed along one or more edges of the sample by scribing through the Ni and the film down to the BN layer. The stressed Ni layer applies a combination of forces which tend to propagate the crack at the 2D layer interface.[24] Thermal release tape is applied to the Ni surface as a handle, and a peeling motion is used to guide the fracture and exfoliate the film. After release, the chosen bonding method can vary depending on the film thickness and secondary substrate characteristics. The film can be transferred and bonded using adhesives before the Ni is removed by wet etching, or if the substrate surface permits, the film can be vdW-bonded before or after the Ni is removed.

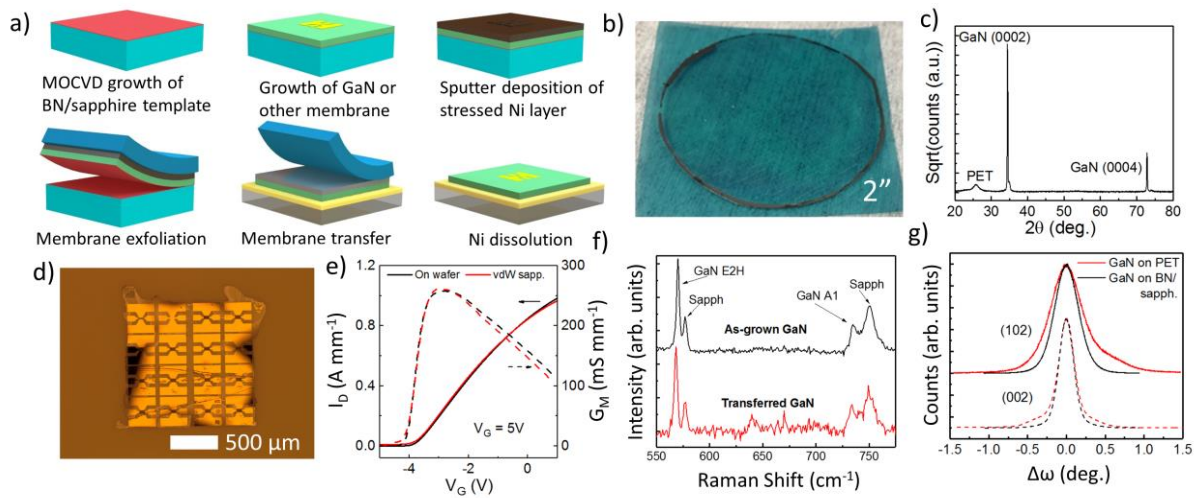

Figure 1: a) Process flow for the Ni-spalling exfoliation and secondary transfer of electronic films. b) Photograph of 2" GaN transferred to PET tape. c) 2θ-ω x-ray scan of the transferred GaN film. d) An array of AlGaN/GaN HEMTs vdW-bonded to a secondary sapphire wafer. e) Transfer curves for a HEMT before and after exfoliation and vdW-bonding to a secondary sapphire wafer. f) Raman spectra of GaN before and after transfer from the sapphire growth wafer to a secondary sapphire wafer adhered using epoxy. g) X-ray rocking curves of the (102) and (002) GaN peaks before and after transfer.

We show that the transfer process is versatile and low impact through extensive pre- and post-transfer thin film and device characterization. A full 2" GaN wafer was easily exfoliated and transferred to a flexible tape as depicted in Figure 1b and 1c while Figure 1d shows an array of HEMTs vdW-bonded to a sapphire substrate. Raman spectroscopy of the GaN films before and after transfer indicates minimal influence on the GaN (Figure 1f), with a small shift in the GaN $E_{2H}$ mode due to strain relaxation (Figure S3).[31] The rocking curves of the GaN (002) and (102) peaks in Figure 1g further demonstrate the minimal influence on thin film properties resulting from the transfer process. Figure 1e shows transfer characteristics of an AlGaN/GaN HEMT with 140nm T-gates before and after transfer and vdW-bonding to a secondary sapphire wafer. Peak transconductance ($G_m$) and current of 260 S mm$^{-1}$ and 1.0 A mm$^{-1}$, respectively, are unchanged as a result of device transfer. Thus demonstrating that the exfoliation process does not damage these delicate devices and that a bond interface with good thermal performance can be achieved due to the atomically smooth surface of the BN layer (AFM results Figure S9).

Exfoliation of the film from the native substrate can allow relaxation of residual film stresses. This, many times, is beneficial for device reliability and even for allowing new heterostructures to be built,[32,5] but it can also present challenges during the transfer process. Depending on the initial film thickness and stress, the Ni layer stress can oppose the film stress such that the film does not completely relax after exfoliation. This behavior is illustrated in **Figure 2a** which depicts the GaN strain in the exfoliated Ni/GaN bilayer as a function of the Ni thickness. As the Ni is then etched away, the GaN would completely relax. However, if the film is rigidly bonded before etching the Ni, the bond interface then becomes stressed and prevents complete relaxation. This situation can cause film cracking if the bond is weak or too flexible or if the residual strain in the exfoliated Ni/GaN stack is large.

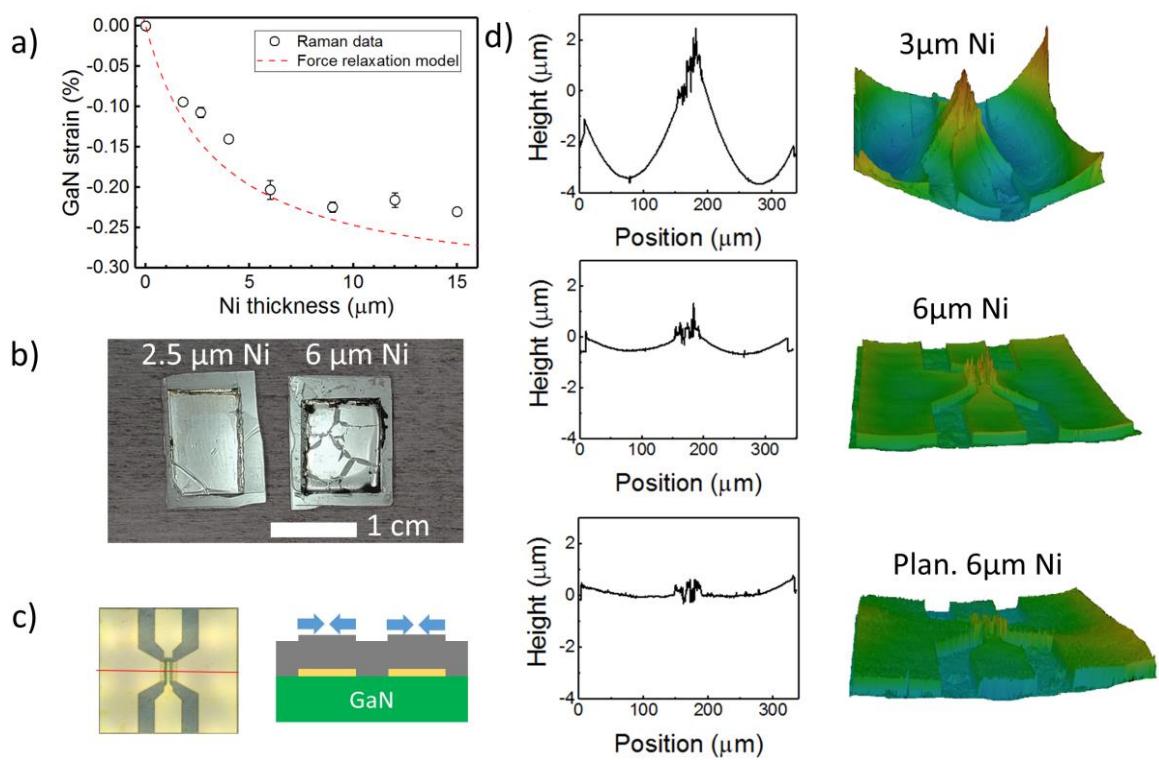

Figure 2: a) GaN strain in the GaN/Ni bilayer as a function of the Ni layer thickness used to exfoliate the film. The data points are derived from the GaN Raman $E_{2H}$ mode frequency, and the red curve is the model. b) Comparison of crack density in GaN membranes transferred to rigid sapphire wafer pieces using different thicknesses of Ni for exfoliation. c) An image of the HEMT showing the orientation of the line scans in d) and a cartoon illustrating the forces which cause the bowing. d) To show the bowing behavior, height line scans and confocal images of transferred HEMTs exfoliated using 3 µm, 6 µm, and planarized 6 µm Ni. The z-scale for all three confocal images is the same.

We analyze and characterize this residual strain issue for exfoliating GaN from BN-on-sapphire templates. Heteroepitaxially grown GaN is generally strained due to lattice and thermal expansion coefficient mismatch with the substrate. In the case of the GaN on BN-on-sapphire films here, the GaN film is measured to have 0.21% compressive strain before separation.[31] After separation from the sapphire substrate, the Ni/GaN bilayer relaxes until the Ni and GaN forces equalize. To understand the relaxation behavior, we measured the GaN $E_{2H}$ Raman mode frequency of a series of bare GaN samples that were exfoliated using different Ni layer thicknesses. Figure 2a shows the GaN strain, derived from the $E_{2H}$ frequency using calculated deformation potentials ($\Delta\omega = (\epsilon_x + \epsilon_x) \times 560$ cm$^{-1}$),[33] as a function of the Ni thickness in the Ni/GaN bilayer (GaN thickness of 2.2 µm). The data point for zero Ni thickness is from a freestanding GaN flake which is assumed to have zero strain. To summarize, as the Ni thickness approaches zero the GaN becomes completely relaxed, while the presence of tensilely stressed Ni opposes the expansion of the compressively strained GaN, reducing the extent of relaxation. The trend in Figure 2a can be more precisely understood using a simple bilayer relaxation model.[23] In this model the GaN and Ni layers each have an initial strain state which is clamped by the sapphire substrate until the bilayer is exfoliated. Upon exfoliation, the bilayer relaxes with the condition that the change in strain upon relaxation is the same for the Ni and GaN layers, i.e. there is no slippage at the Ni/GaN interface. The condition for equilibrium is that the forces equalize.

$$t_{Ni}C_{Ni}(\epsilon_{Ni}^0 + \Delta\epsilon) + t_{GaN}C_{GaN}(\epsilon_{GaN}^0 + \Delta\epsilon) = 0 \qquad (1)$$

$t_i$ is the layer thickness, $C_i$ is the layer elastic modulus, $\epsilon^0_i$ is the initial layer strain, and $\Delta\epsilon$ is the change in strain upon relaxation. The GaN strain as a function of the Ni thickness can then be expressed as the following (See the SI for details about the parameters in the model):

$$\epsilon_{GaN}(t_{Ni}) = \epsilon_{GaN}^0 + \frac{-t_{Ni}C_{Ni}\epsilon_{Ni}^0 - t_{GaN}C_{GaN}\epsilon_{GaN}^0}{t_{Ni}C_{Ni} + t_{GaN}C_{GaN}} \qquad (2)$$

  The data in Figure 2a shows there is a significant difference in exfoliated GaN strain depending on the thickness of the stressed Ni layer used. This residual strain tends to cause cracks during the transfer process especially when the GaN is bonded using an adhesive before the Ni is etched away. In this situation, as the Ni is etched, the stress at the GaN/adhesive interface increases, and if the adhesive bond is not strong enough the GaN film buckles and cracks. Figure 2b shows a qualitative comparison in crack density between using thick (6 µm) and thin (2.5 µm) Ni to exfoliate the GaN. These ~1 cm membranes were adhered to sapphire substrates with epoxy before the Ni was etched away. The GaN which was exfoliated using the thicker 6 µm Ni had more residual strain which resulted in more cracking.
  In addition to the residual in-plane strain effects, we observed significant bowing effects from the stressed Ni layer when exfoliating processed HEMTs. The strain gradient set up by the stressed Ni layer in part provides the energy for exfoliation, but it also causes the exfoliated film to bow and exhibit substantial curvature. For a homogeneous film, the induced curvature is uniform, but for devices comprised of numerous materials with inhomogeneous thickness and mechanical properties, the curvature can be inhomogeneous and localized. Upon exfoliation of AlGaN/GaN HEMTs we observed localized bowing where each isolated contact pad was distorted into a dish-like shape. The bowing was present as soon as the GaN was separated from the substrate. Bonding the samples to a secondary sapphire substrate using an epoxy layer preserved the bowed shape even after the Ni was etched away. Confocal images were taken to characterize the vertical bowing in these devices and are depicted in Figure 2d. To understand the local bowing behavior in the contact region, we consider the stress applied by the Ni layer on different device regions. The contact regions include a 500 nm thick metal stack which represents a significant inhomogeneity in the total Ni/GaN device layer thickness. The stressed Ni applies a bending moment relative to the neutral bending plane of the Ni/GaN bilayer. Since the Ni deposition is conformal, the contact regions put the Ni layer 500 nm farther away from the neutral plane, so it applies a greater bending moment. This situation causes the concave-up bowing of the contact regions. We observed that exfoliated films with thicker 6 µm Ni exhibited less bowing which is consistent with our understanding since using thicker Ni reduces the relative difference in applied moment between on and off the contact region. To further address the bowing issue, we planarized the top Ni surface prior to exfoliation. Polishing the top Ni surface until the contact areas were level with the rest of the surface eliminated the spatial variations in bending moment associated with the contact stack height. In Figure 2d, confocal images reveal the differing topography and resultant height line scans of HEMTs transferred using 3 µm, 6 µm, and 6 µm planarized Ni. For the non-planarized transfers, each contact pad is bowed up which causes a large curvature in the channel region of the HEMT and a significantly increased chance of failure and degradation. As can be seen in Figure 2d, the planarization appeared to eliminate the curvature in the channel region.

In addition to large areas, this 2D-layer-mediated Ni-spalling technique is conducive to small and precise selective area transfer. Scribing through the Ni and GaN layers down to the BN initiates the crack, and depending on the Ni stress and thickness, the crack will propagate spontaneously or a handle can be attached to apply the required bending moment. See Figure S4 for examples. This crack initiation step can be used to define a selected area for separation by outlining or drawing a crack with a laser milling tool or mechanical scribe. **Figure 3a** shows the process flow for the selective area separation, and photographs of single HEMTs outlined and separated using this process in Figure 3b and 3c. Recently, selective area liftoff of GaN films grown on BN was demonstrated but required masking with $SiO_2$ prior to BN film deposition.[34] The back-end selective liftoff process described herein maximizes usable wafer area and limits the number of alignment steps required. Furthermore, the processed device layout or alignment can be maintained during the transfer process which could be used to build multiple circuits simultaneously, saving additional time and alignment steps. See Figure S7 for photographs of aligned devices during transfer. The process is also amenable to transferring a HEMT to a flexible substrate. Figure 3d shows transfer curves for the HEMT with substrate flat and while bent, inducing 0.05% uniaxial strain in the GaN, as derived from the $E_{2H}$ frequency shift. The HEMT showed no measurable change in DC performance with the substrate flexed. The selective area transfer strategy can effectively utilize large densely organized processed wafers and represents an important step towards the use of thin film transfer in new heterogeneous integration designs.

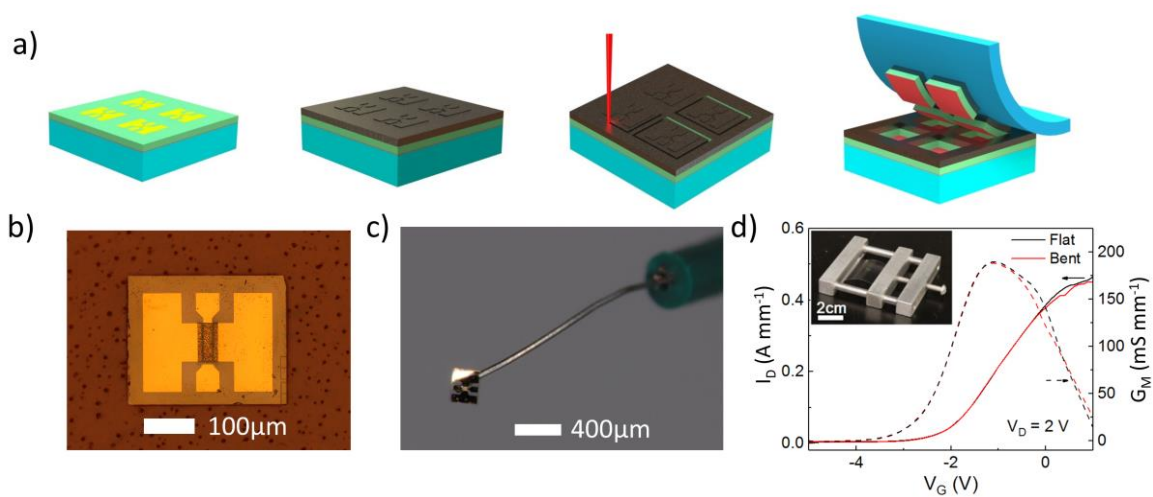

Figure 3: a) Process flow for selective area exfoliation. b.) A single HEMT transferred to Kapton tape. c) A single HEMT on a strand of wire. d) Transfer curve of a HEMT transferred to a flexible PET substrate flat and while bent and strained. In the inset, a photograph of the bending apparatus.

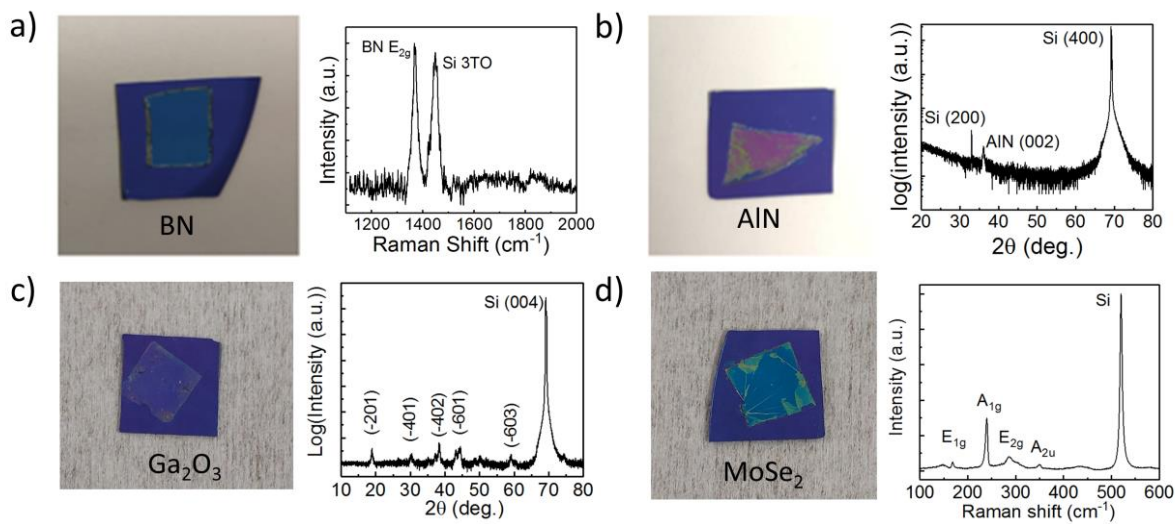

Figure 4: Additional electronic films transferred using the Ni-spalling technique. Films are approximately 1 cm in size. a) sp$^2$-bonded BN film transferred to SiO$_2$/Si substrate. Raman spectroscopy confirms the presence of BN. b) 20 nm AlN film transferred to SiO$_2$/Si substrate. X-ray diffraction data showing the AlN 002 peak. c) 100 nm Ga$_2$O$_3$ film vdW-bonded to SiO$_2$/Si wafer piece. X-ray data confirming the presence of Ga$_2$O$_3$. d) MoSe$_2$ film transferred to SiO$_2$/Si substrate. Raman spectroscopy confirms its presence.

The robustness of the BN/sapphire template allows this technique to be adapted to many different semiconducting and insulating layers beyond the already demonstrated GaN films and devices. The BN layer grown by MOCVD is more stable than graphene and TMDs in inert and oxidizing atmospheres,[35-37] which enables growth and processing of more films with harsher growth conditions while still allowing easy exfoliation after film processing. Here we demonstrate transfer of a variety of films which could be used to provide new device functionality or improve device performance. The BN layer itself can be transferred to function as an insulating or passivating layer.[37,38] **Figure 4a** shows a sp$^2$-bonded BN film transferred to a SiO$_2$/Si substrate, with the Raman spectrum confirming the transfer of the BN. MOCVD-grown AlN and Ga$_2$O$_3$ films were also transferred to a SiO$_2$/Si substrate, shown in Figure 4b and 4c. AlN films have been receiving interest for their high thermal conductivity for heat spreading applications. The ability to transfer high quality AlN ultra-thin films could be an important strategy for improving heat dissipation in flexible electronics, where typical AlN growth temperatures are too high for many flexible substrates.[39] Ga$_2$O$_3$ is a promising material for power conversion because of its high breakdown field strength and temperature stability, but its low thermal conductivity might be a performance bottleneck. Transferring the transistors on just the thin semiconductor membrane to a high thermal conductivity substrate could improve their performance.[40] Finally, we transferred a MoSe$_2$ film using the vdW-mediated spalling method. The ability to precisely transfer TMDs and other semiconducting 2D materials opens new possibilities for high performance and ultra-efficient devices.[40] See the SI for additional films exfoliated.

## 3. Conclusion

New heterogeneous integration strategies are likely to be needed as smaller and lighter communication platforms are introduced. Isolating and integrating just the thin active regions of electronic components has the potential to maximize thermal performance, size and weight savings, and even allow for more functionality. Herein, we have presented a membrane transfer method which allows for exfoliation of thin electronic layers from BN-on-sapphire templates using a stressed metal layer. We explored some of the challenges associated with transferring strained membranes such as cracking and bowing and have presented strategies to alleviate those issues. This versatile process reveals potential significant improvements to other transfer techniques including the robustness to post-processing and ability to separate both wafer-scale and precise small device areas. Together these abilities can result in new heterogeneous integration strategies of great interest to the electronics communities.

## 4. Experimental Section/Methods

*Boron nitride growth*: All BN-on-sapphire templates used in this work were grown on (0001) sapphire by MOCVD using triethylborane (TEB) and ammonia ($NH_3$) at 1050 °C, 20 Torr, and V/III ratio of 1800 to 2700 for 30 minutes. Under these conditions the growth is self-terminating and produces highly reproducible $sp^2$ BN films with a thickness of 1.6 nm and RMS roughness ~0.1 nm.[30]

*III-Nitride growth:* AlN, GaN and AlGaN/GaN layers were grown on BN-on-sapphire templates in an EMCORE D180 turbo disk MOCVD using ammonia, trimethylaluminum (TMA), trimethylgallium (TMG), and ferrocene as precursors for N, Al, Ga, and Fe, respectively. The thin (~20 nm) AlN layers in Figure 4 were deposited at 1000 °C and 60 Torr using a V/III ratio of 2300. For GaN and AlGaN/GaN films a ~15 nm thick AlN nucleation layer was deposited followed by a standard three step GaN process.[31] For AlGaN/GaN HEMT structures, a 1.6 µm Fe-doped GaN, 0.4 µm of undoped GaN, ~2 nm AlN insert layer, 17 nm AlGaN boundary layer with 27% Al and 3 nm GaN cap layer were grown.

*$Ga_2O_3$ growth:* β-$Ga_2O_3$ thin films were grown on BN-on-sapphire templates by MOCVD. The $Ga_2O_3$ growth was performed in a vertical quartz reactor from Agnitron Agilis designed with a far-injection showerhead.[41] Triethygallium (TEGa) and oxygen gas were used as the Ga and O precursors and argon as the carrier gas. The un-doped $Ga_2O_3$ growth was performed at a temperature of 600 °C and a chamber pressure of 60 Torr. Post-growth XRD measurements show that (-201) oriented β-$Ga_2O_3$ films grew on the BN-on-sapphire templates.

*$MoSe_2$ deposition:* $MoSe_2$ thin films were deposited on BN-on-sapphire templates using a two-step process. First, a 0.7 nm thick Mo layer was deposited by pulsed DC sputtering using an Advanced Energy Pinnacle Plus power supply at 90 W and a pressure of 10 mTorr Ar. Mo films on BN/sapphire were then transferred into a CVD reactor evacuated and purged with a flow of $H_2$. For selenization, the films were heated to 650 °C under a flow of $N_2/H_2$ (95%:5%). After reaching 650 °C, films were selenided under a flow of $H_2Se$ for 30 minutes then cooled to 400 °C before turning off the flow of $H_2Se$.

*HEMT fabrication:* High electron mobility transistors used in this work were fabricated from AlGaN/GaN structures on BN/sapphire templates prior to liftoff similar to the process described in ref. [43]. Isolation mesas were etched using inductively coupled plasma with $BCl_3/Cl_2$/Ar chemistry to a depth of ~80 nm. Then Ti/Al/Ni/Au Ohmic contacts were deposited by electron beam evaporation and annealed at 850 °C for 30 seconds in nitrogen ambient using a rapid thermal anneal (RTA). Next, 0.14 µm Ni/Au Schottky T-gates were formed using electron beam lithography. Then Ti/Au metal interconnects and 200 nm $Si_3N_4$ passivation layers were deposited.

*Stressed Ni deposition:* Ni was deposited using RF magnetron sputtering. Prior to Ni deposition, films and device surfaces were treated with a 200W oxygen plasma for two minutes. For GaN films, a 30 nm thick Cr adhesion layer was deposited first by evaporation. The Ni was sputtered at 300 W with 5 mTorr of Ar, and the deposition rate was approximately 2 µm h$^{-1}$. The Ni stress was measured to be on average 0.27 GPa using the curvature of a Si wafer. See the SI for more Ni sputtering details.

*GaN film and HEMT transfer:* After Ni sputtering, to initiate controlled spalling, a crack is initiated by scribing through the Ni and GaN down to the BN layer using a diamond scribe or a laser milling tool. Thermal release tape is applied to the Ni surface and the crack is propagated by applying a peeling motion to the tape. For the transfers depicted in Figure 1b and 2b, the GaN was bonded using adhesives before the thermal release tape was removed by heating on a hotplate. Then the Ni was removed by dripping Transene Ni TFB etchant on the Ni surface. Depending on the compatibility of the adhesive, the entire sample can be dipped in the etchant. After the Ni is removed, the surface is rinsed with DI water and the thin Cr layer is etched using Transene Cr etchant.

For the vdW-bonded HEMTs and selectively transferred HEMTs in Figure 1d and 3b, respectively, the Ni was instead removed prior to bonding. The selectively separated HEMT was outlined using a 3D-Micromac MicroPREP laser milling tool at 0.3 W before being separated with thermal release tape. The devices were released from the thermal release tape after separation from the growth substrate, then in a bath of liquid the metals were etched from the freestanding GaN/device flakes. Using this method, the residual strain in the GaN is completely relaxed before bonding. For the devices in Figure 1d, the flake was vdW-bonded to a sapphire substrate using water surface tension-mediated bonding. For the single HEMT in Figure 3b, the HEMT was picked up using a PDMS stamp then bonded to the Kapton tape.

*Other film transfers:* The thin films vdW-bonded in Figure 4 were transferred using a PMMA float transfer method. After approximately 1.5 µm of Ni was sputtered, the films were separated using thermal release tape. PMMA was spin coated on the backside of the films to a thickness of approximately 350 nm then was baked at 70 ˚C for 10 minutes. The films were then released from the thermal release tape and floated Ni side down on Ni etchant. After the Ni was removed, the films were transferred to a beaker of DI water before being float-transferred to the SiO$_2$/Si substrates. The PMMA was dissolved using acetone after leaving the film to dry for approximately 24 hours.

*Raman spectroscopy:* A Renishaw Inviva microscope was used to collect the Raman data. For the GaN and MoSe$_2$ Raman a 532 nm diode laser was used. A Si wafer was used to calibrate the system for the strain measurements in Figure 2a. The BN Raman spectrum in Figure 4a was taken using a 488 nm laser.

*X-ray data:* X-ray scans were collected using a Malvern Panalytical Empyrean system.

*Electrical measurements:* The transistor transfer curves in Figure 1 and 3 were taken using a Keithley 2636 source measure unit with custom LabView software.

**Supporting Information**

Supporting Information is available from the author.

**Acknowledgements**

This work was funded by the Air Force Office of Scientific Research under Award No. FA9550-19RYCOR050.